\documentclass[aps,twocolumn,prb,preprintnumbers,showpacs,amsmath,amssymb,superscriptaddress]{revtex4-2}

\usepackage{epsfig}
\usepackage{graphicx}
\usepackage{dcolumn}
\usepackage{bm}
\usepackage{tabularx}
\usepackage[pdftex, colorlinks, citecolor=blue]{hyperref}   
\usepackage{appendix}
\usepackage{diagbox}
\usepackage{threeparttable}
\usepackage{multirow}
\usepackage{booktabs}
\usepackage{txfonts}
\usepackage{xcolor}
\usepackage{amssymb}
\usepackage{amsmath}
\usepackage{latexsym}
\usepackage{amsbsy}
\usepackage{array}
\usepackage{esvect} 
\usepackage{extarrows}
\usepackage{float}
\usepackage[british]{babel}
\usepackage{upgreek}

\newcommand{\bl}[1]{\textcolor{blue}{#1}}

\begin{document}

\title{Enhanced superconductivity in atomically thin noble metals: From quantum confinement to interface-induced Lifshitz transition}

\author{Chun-Jie Zhang}
\thanks{These authors contributed equally to this work.}
\affiliation{School of Physical Science and Technology, Inner Mongolia University, Hohhot, 010021, China
}

\author{Bing Zhang}
\thanks{These authors contributed equally to this work.}
\affiliation{School of Physical Science and Technology, Inner Mongolia University, Hohhot, 010021, China
}

\author{Yapeng Wu}
\affiliation{School of Physical Science and Technology, Inner Mongolia University, Hohhot, 010021, China
}

\author{Xiao-Ping Li}
\affiliation{School of Physical Science and Technology, Inner Mongolia University, Hohhot, 010021, China
}
\affiliation{Key Laboratory of Semiconductor Photovoltaic Technology and Energy Materials at Universities of Inner Mongolia Autonomous Region,
Inner Mongolia University, Hohhot 010021, China
}
\affiliation{Research Center for Quantum Physics and Technologies, Inner Mongolia University, Hohhot 010021, China
}
\author{Lei Wang}
\email{lwang@imu.edu.cn}
\affiliation{School of Physical Science and Technology, Inner Mongolia University, Hohhot, 010021, China
}
\affiliation{Research Center for Quantum Physics and Technologies, Inner Mongolia University, Hohhot 010021, China
}
\affiliation{Inner Mongolia Key Laboratory of Microscale Physics and Atomic Manufacturing, Inner Mongolia University, Hohhot 010021, China
}
\begin{abstract}
Unlocking superconductivity in intrinsically non-superconducting noble metals Au, Ag and Cu represents a fundamental challenge in low-dimensional quantum materials. While quantum confinement in the atomically thin limit is known to trigger emergent superconductivity, strategies to amplify this marginal effect to experimentally accessible temperatures remain a key open question. Using first-principles calculations, we establish a unified framework linking intrinsic confinement effects with interface engineering in noble metal films. We reveal that intrinsic superconductivity is element-specific: it is suppressed in Ag by a stiff phonon spectrum, but emerges in trilayer Cu with $T_\text{C} \approx 0.78$ K and pentalayer Au with $0.63$ K driven by confinement-induced density of states (DOS) enhancement and phonon softening, respectively. In h-BN/Cu(111) heterostructures, $T_\text{C}$ is critically by the interfacial stacking configuration. The thermodynamically stable N-bonded interface provides an experimentally accessible superconducting state with $T_\text{C} \approx 3.23$ K, whereas a metastable B-bonded configuration boosts $T_\text{C}$ to 7.00 K. Detailed analysis reveals that the enhancement is not caused by charge transfer or a simple increase in the DOS, but by an interface-induced Lifshitz transition that reshapes
the Fermi surface and amplifies momentum-resolved electron--phonon coupling matrix elements. This Lifshitz-controlled mechanism persists in bilayer h-BN/3L-Cu(111) and is further reproduced in heterometallic 3L-Au/Ag(111), highlighting its transferability beyond a specific interface configuration. Our work unifies the understanding of intrinsic two-dimensional superconductivity with atomistic interface design, and identifies interface-controlled Fermi-surface topology as a route to functionalizing noble metals as emergent superconductors.
\end{abstract}

\maketitle

\section{INTRODUCTION}
Bulk noble metals—gold (Au), silver (Ag), and copper (Cu)—constitute the bedrock of modern electronics and plasmonics  \cite{re1,re2,re3,re4,re5,re6,re7}, celebrated for their superior conductivity and chemical stability  \cite{re8,re9,re10}. Despite their prevalence as archetypal Fermi liquids, they present a fundamental dichotomy in condensed matter physics: while they possess high carrier densities, they lack the requisite electron-phonon instability to form Cooper pairs, rendering them nonsuperconducting down to the lowest accessible temperatures  \cite{re11,re12,re13}. Consequently, inducing an intrinsic superconducting state in these elemental metals remains an elusive goal. Realizing this would not only address the dissipation limits in cryogenic quantum devices \cite{re14,re15,re16,re17,re18,re19} but also establish a pristine, chemically simple platform for investigating low-dimensional quantum coherence \cite{re20,re21}.

Over the past decades, the primary route to inducing superconducting states in these materials has been the proximity effect within heterostructures that incorporate conventional superconductors, where phase coherence is borrowed from an external pairing reservoir. While crucial for mesoscopic devices, this approach does not confer superconductivity as an intrinsic property of the noble metal itself \cite{re22}. The framework of quantum confinement offers a compelling theoretical pathway to bridge this gap \cite{re23,re24}. In the atomically thin limit, the formation of quantum well states discretizes the electronic band structure, generating van Hove singularities that can theoretically enhance the density of states (DOS) at the Fermi level \cite{re25,re26,re27,re28,re29}. However, translating this theoretical potential into robust superconductivity faces fundamental hurdles. Intrinsic noble metal films often suffer from the absence of soft phonon modes critical for strong pairing, which counteracts the benefits of DOS enhancement, resulting in weak electron–phonon coupling (EPC) and vanishingly low transition temperatures ($T_\text{C}$) \cite{re30}. Moreover, theoretical predictions often rely on free-standing approximations that neglect environmental screening and interfacial hybridization, limiting their predictive power for realistic experimental setups \cite{re31,re32}. Thus, a critical hurdle remains: how to actively bypass this intrinsic stiff phonon bottleneck to engineer robust pairing in noble metal films.

Interface engineering in van der Waals (vdW) heterostructures provides the necessary degrees of freedom to transcend these limitations \cite{re33,re34}. Beyond simple encapsulation, heterointerfaces introduce proximity effects, charge transfer, and orbital symmetry breaking  \cite{re35,re36,re37,re38,re39}. These phenomena can fundamentally renormalize the superconducting order parameter, as demonstrated in systems like IrTe$_2$/In$_2$Se$_3$ \cite{re40,re41} and 1T'-WTe$_2$/2H-NbSe$_2$ \cite{re42}. In this context, noble-metal/hexagonal boron nitride (h-BN) heterostructures represent a promising yet underexplored frontier. The atomic flatness and wide bandgap of h-BN facilitate the formation of atomically sharp interfaces \cite{re43,re44,re45}, and recent advances in synthesizing single-crystal h-BN/Cu(111) films \cite{re46} bring these systems within experimental reach. A pivotal question emerges: Can the h-BN interface act as more than a passive substrate? Specifically, can precise interfacial orbital engineering synergize with quantum confinement to amplify the electron-phonon interaction, thereby unlocking significant superconductivity in noble metals?

In this work, we present a systematic first-principles study to address these questions, elucidating the microscopic mechanisms governing superconductivity in ultrathin noble metal films (Au, Ag, Cu) and their h-BN-functionalized interfaces. We first uncover the thickness-dependent evolution of $T_\text{C}$, disentangling the competition between confinement-induced DOS enhancement and lattice stiffness to identify the material-specific constraints. Building on this, we demonstrate that interface engineering in h-BN/Cu(111) offers a powerful control knob. By analyzing a hierarchy of configurations—from the thermodynamically stable monolayer interface to slip-stacked bilayers and confined sandwich structures \cite{re47}—which serve to probe the operational limits—we reveal that the enhancement is driven by specific interfacial stacking configuration and interface-induced Lifshitz transition. Our results provide a unified physical picture, demonstrating how atomistic interface design can elevate a marginal superconductor into a robust one, offering a blueprint for functionalizing noble metals in the quantum regime.

\section{COMPUTATIONAL METHODS}
First-principles calculations were performed within the framework of density functional theory (DFT) as implemented in the QUANTUM ESPRESSO package \cite{re48}. The exchange-correlation interaction was described using the generalized gradient approximation (GGA) parameterized by Perdew, Burke, and Ernzerhof (PBE) \cite{re49,re50}. Ion-electron interactions were modeled using norm-conserving pseudopotentials \cite{re51}. The kinetic energy cutoffs for wave functions and charge density were set to 90 Ry and 360 Ry, respectively. For Brillouin zone integration, a $24\times24\times1$ Monkhorst-Pack $k$-point grid was employed, combined with a Methfessel-Paxton smearing width of 0.02 Ry to ensure convergence. A 25 \AA~vacuum layer was applied along the out-of-plane direction to eliminate spurious interactions between periodic images. Structural relaxations were performed until the total energy change was less than $10^{-6}$ Ry and the residual forces on each atom fell below $10^{-6}$ Ry/Bohr (Note S1). Phonon dispersion and electron-phonon coupling (EPC) properties were calculated using density functional perturbation theory (DFPT) \cite{re52}.The dynamical matrices were computed on an $8\times8\times1$ $q$-point grid. For the h-BN/Cu(111) heterostructures, van der Waals (vdW) interactions were accounted for using the DFT-D2 semi-empirical correction scheme \cite{re53} (additional functionals tested; see Figure S8 and Table S3). The EPC matrix elements $g_{nm,\nu}(k,q)$ were first computed \cite{re54,re55,re56,re57} based on the above $\bm{k}$ and $\bm{q}$ grids, where $m$ and $n$ are band indices, and $\nu$ indicates a phonon branch. The matrix elements quantify the scattering amplitude between the electronic states with a wave vector $\boldsymbol{k}$, a band index $m$ (denoted as $(\boldsymbol{k}, m)$), and $(\boldsymbol{k} + \boldsymbol{q}, n)$ through a phonon mode with a branch $\nu$ and a wave vector $\boldsymbol{q}$.

The electron–phonon interaction strength was obtained via the Migdal-Eliashberg formalism \cite{re58,re59}:
\begin{equation}
\lambda_{q\nu} = \frac{\gamma_{q\nu}}{\pi N(\varepsilon_F) \omega_{{q}\nu}^2},
\end{equation}
where $\gamma_{\bm{q}\nu}$ is the phonon linewidth, $\omega_{\bm{q}\nu}$ is the phonon frequency, and $N(\varepsilon_F)$ is the electronic DOS at the Fermi level. The $\gamma_{\bm{q}\nu}$ is defined by:
\begin{equation}
\gamma_{\bm{q}v}=\frac{2\pi{\ \omega}_{\bm{q}\nu}}{\mathrm{\Omega}_{BZ}}\ \ \sum_{\bm{k},n,m}{\mid g_{nm,v}(k,q)\mid^2\delta(\varepsilon_{nk}-\varepsilon_F)\delta(\varepsilon_{mk+q}-\varepsilon_F)}
\end{equation}
where $\mathrm{\Omega}_{\mathrm{BZ}}$ is the volume of the Brillouin zone, and $\varepsilon_{nk}\ (\varepsilon_{m(k+q)})$ are the Kohn-Sham eigenvalues. The Eliashberg electron-phonon spectral function $\alpha^2F(\omega)$ is then calculated by:

\begin{equation}
\alpha^2F(\omega) = \frac{1}{2\pi N(\varepsilon_F)} \sum_{\bm{q}\nu} \frac{\gamma_{\bm{q}\nu}}{\hbar \omega_{\bm{q}\nu}} \delta\left( \omega - \omega_{\bm{q}\nu} \right).
\end{equation}

The total EPC $\mathrm{\lambda}$ can be calculated by integrating the Eliashberg spectral function \cite{re60}:
\begin{equation}
\lambda(\omega)=\sum_{\bm{q}\nu}{\lambda_{\bm{q}\nu}=\mathrm{2} \int_{0}^{\infty}\frac{\alpha^2F(\omega)}{\omega}d\omega\ }
\end{equation}
The superconducting critical temperature $T_\text{C}$ was estimated using the McMillan-Allen-Dynes formula \cite{re61,re62}:
\begin{equation}
T_{C} = \frac{\omega_{\mathrm{log}}}{1.2} \exp \left [ \frac{-1.04(1+\textnormal{$\lambda$})}{\textnormal{$\lambda$} -u^{*}(1+0.62\textnormal{$\lambda$})} \right ] ,
\end{equation}

where $\mu^*$ is the effective screened Coulomb repulsion constant, set to 0.11 \bl{\cite{re63,re64}} (a commonly adopted value for weakly correlated systems). The logarithmic average phonon frequency $\omega_{\text{log}}$ was calculated via:
\begin{equation}
\omega_{\text{log}} = \exp\left( \frac{2}{\lambda} \int_0^\infty \frac{\alpha^2F(\omega)}{\omega} \log\omega \, d\omega \right).
\end{equation}
All crystal structures presented in this work are visualized using the VESTA software \cite{re65}.

\section{RESULTS AND DISCUSSION}

\subsection{Lattice structures}
\begin{figure*}[htbp]
\centering
\includegraphics[width=1\textwidth]{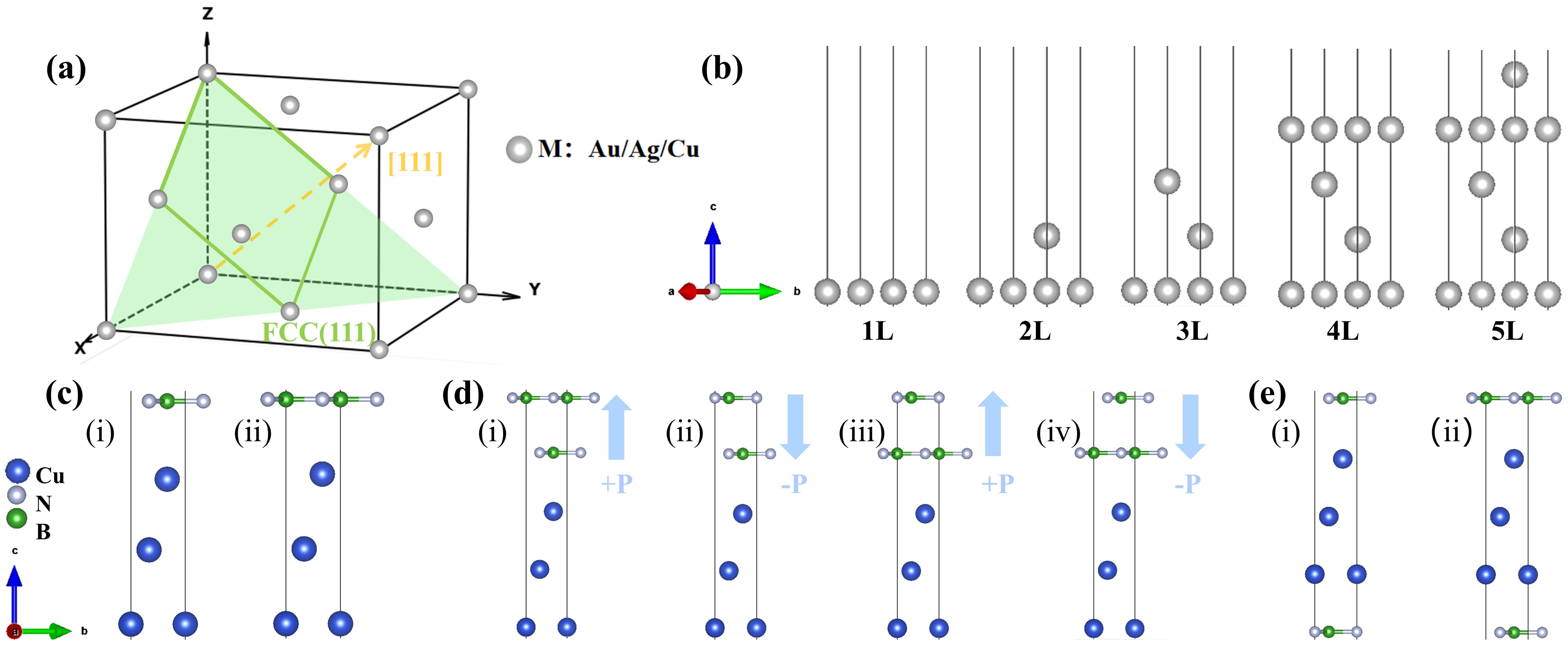} 
\caption{Crystal structures. (a) FCC lattice with the (111) plane and [111] direction. (b) ABC-stacked M (M = Au, Ag, Cu)films with 1 to 5 layers. (c) Side views of h-BN/3L-Cu(111) heterostructures with the interface at the (i) B-site and (ii) N-site. (d) Side views of h-BN/h-BN/3L-Cu(111) heterostructures with AB/BA-stacked (+P/-P) bilayer h-BN on the (i,~ii) B-site and (iii,~iv) N-site, respectively. (e) Side views of h-BN/3L-Cu(111)/h-BN sandwich heterostructures with the interface at the (i) B-site and (ii) N-site.}
\label{fig1}
\end{figure*}

\begin{ruledtabular}
\renewcommand\arraystretch{1.2}
\begin{table*}
\caption{Structural parameters of noble metal films. Bulk values, optimized in-plane lattice constants, and thicknesses for Ag, Cu, and Au films with 1–7 layers.}
\begin{tabular}{cccccccccccc}
& \multicolumn{3}{c}{Ag (2.889 \AA)} & & \multicolumn{3}{c}{Cu (2.556 \AA)} & & \multicolumn{3}{c}{Au (2.884 \AA)} \\ 
\cline{2-4} \cline{6-8} \cline{10-12}
& Optimized & Thickness & Interlayer & & Optimized & Thickness & Interlayer & & Optimized & Thickness & Interlayer \\
& (\AA)     & (\AA)     & spacing(\AA)& & (\AA)     & (\AA)     & spacing(\AA)& & (\AA)     & (\AA)     & spacing(\AA) \\
\cmidrule(lr){1-12}
1L & 2.79 & --     & -- &    & 2.43     & --     & -- &    & 2.74     & --     & --     \\
2L & 2.85 & 2.47 & 2.47 &    & 2.50 & 2.11 & 2.11 &  & 2.77 & 2.86 & 2.84 \\
3L & 2.89 & 4.83 & 2.42 &    & 2.52 & 4.21 & 2.12 &  & 2.82 & 5.18 & 2.57 \\
4L & 2.90 & 7.19 & 2.40 &    & 2.54 & 6.29 & 2.10 &  & 2.84 & 7.64 & 2.55 \\
5L & 2.91 & 9.65 & 2.41 &    & 2.54 & 8.41 & 2.10 &  & 2.86 & 10.06 & 2.51 \\
6L & 2.92 & 11.96 & 2.39 &    & 2.55 & 10.49 & 2.10 &  & 2.87 & 12.48 & 2.50 \\
7L & 2.91 & 14.39 & 2.40 &    & 2.55 & 12.60 & 2.10 &  & 2.88 & 14.90 & 2.48
\label{table1}
\end{tabular}
\end{table*}
\end{ruledtabular} 

To elucidate quantum confinement effects in face-centered cubic (FCC) noble metals (Au, Ag, and Cu), we constructed free-standing ultrathin films derived from the FCC(111) lattice with thicknesses of 1--7 atomic layers, corresponding to the close-packed plane with the lowest surface energy [Figure 1(a) and Figure S1]. Figure \ref{fig1}.~(b) highlights the thin films with 1-5 atomic layers, where quantum confinement effects are expected to be most pronounced. The 1‑ and 2‑layer films represent the ultrathin limit. The 3‑layer film (which forms the first complete ABC stacking unit) and the 5‑layer film serve as key benchmarks for strong quantum confinement, reflecting extremal behaviors in $T_\text{C}$ \cite{re24}. The 6‑ and 7‑layer films capture the gradual convergence toward bulk‑like properties. Our structural optimization reveals a consistent lattice relaxation trend across all three noble metals, as summarized in Table \ref{table1}. The in-plane lattice constant exhibits a contraction effect in the monolayer limit and monotonically increases toward the bulk value with increasing thickness. More significantly, the interlayer spacing ($d_z$) demonstrates a distinct element-specific evolution driven by surface relaxation. In Au films, $d_z$ contracts from 2.84 \AA~in the bilayer (2L) to 2.48 \AA~in the septuple-layer (7L) film. Ag exhibits a similar but milder trend, relaxing from 2.47 \AA~(2L) to 2.40~\AA~(7L). In contrast, Cu displays negligible variation, maintaining a nearly constant spacing of ~2.10 \AA~from the bilayer onward. Remarkably, for Au and Ag, these equilibrated spacings remain appreciably expanded compared to their theoretical bulk (111) interplanar distances (Au: 2.36 \AA, Ag: 2.36 \AA), whereas Cu closely retains its bulk-like spacing (2.08 \AA). This thickness-dependent structural evolution—particularly the anomalous expansion in heavy noble metals—provides the physical groundwork for understanding the modulation of electronic DOS and phonon stiffness that ultimately governs $T_\text{C}$. Finally, to probe interfacial coupling effects, we constructed three prototypical heterostructures based on the 3L-Cu(111) film. First, a monolayer h-BN/Cu(111) system was modeled [Figure \ref{fig1}.~(c)], motivated by the prevalent use of Cu(111) films as catalytic substrates for high-quality h-BN growth \cite{re46,re66,re67}. Second, to explore the universality of interfacial control and the potential role of ferroelectric polarization, we designed a heterostructure comprising a switchable, slip-stacked AB and BA h-BN bilayer on 3L-Cu(111) [Figure \ref{fig1}.~(d)]. Third, we constructed a symmetric h-BN/3L-Cu(111)/h-BN sandwich configuration [Figure \ref{fig1}.~(e)]. This structure serves as a critical control model, allowing us to isolate the effects of interfacial symmetry breaking and distinguish between single-interface hybridization and dual-interface confinement effects.

\subsection{Properties of noble films}
\subsubsection{\textbf{Electronic structure}}
\begin{figure*}[ht]
\centering
\includegraphics[width=1\textwidth]{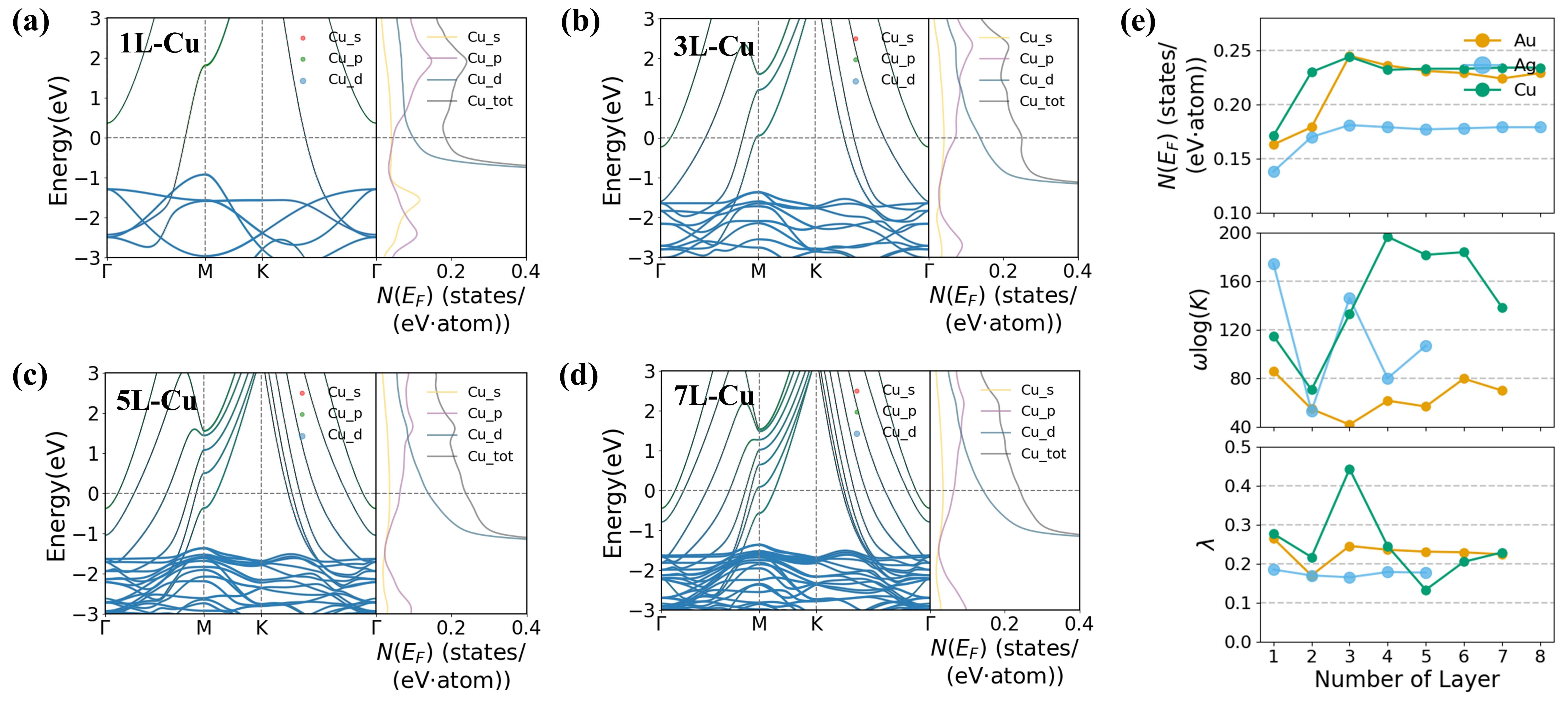}

\caption{Thickness-dependent electronic and superconducting properties. (a-d) Band structures and orbital-resolved DOS of Cu thin films with different thicknesses: monolayer (1L), trilayer (3L), pentalayer (5L), and heptalayer (7L). (e) Superconducting properties of Au, Ag and Cu thin films as a function of layer number: DOS at the Fermi level $N(\varepsilon_\text{F})$, the logarithmic average phonon frequency $\omega_\text{log}$, and the electron-phonon coupling constant $\lambda$.}
\label{fig2}
\end{figure*}

Reducing the thickness of noble metal films (Au, Ag, Cu) to the atomic limit dramatically enhances quantum confinement along the out-of-plane ($z$) direction, triggering a dimensional crossover in their electronic structure from bulk-like to strictly two-dimensional [Figure~S2]. Despite element-specific differences in $d$-band centers and bandwidths, all three systems exhibit a convergent evolution driven by the formation of discrete quantum well states.~These states are characterized by Fermi-level band crossings and strong in-plane delocalization, evidenced by pronounced band dispersion along the M–K path. The electronic structures of our multilayer Au, Ag, and Cu films are validated against literature results \cite{re68}, confirming the dimensional crossover (Note S1). As the film thickness increases, the number of subbands near the $\Gamma$ point grows monotonically. Crucially, the DOS at the Fermi level $N(\varepsilon_\text{F})$ displays a non-monotonic dependence: it peaks at the trilayer limit and saturates as the film approaches bulk-like thickness [Figure \ref{fig2}].

To elucidate the microscopic origin of this evolution, we take Cu thin films as a representative prototype [Figure \ref{fig2}.~(a–d)]. In the monolayer limit [Figure \ref{fig2}.~(a)], the extreme quantum confinement---arising from the complete absence of interlayer coupling---results in bands with large curvature and steep dispersion crossing the Fermi level, indicative of a small effective electron mass.~While the $d$-band DOS exhibits a sharp onset at approximately $-0.50$ eV, the Fermi-level DOS remains relatively low at $N(\varepsilon_\text{F}) = 0.19$ states/(eV$\cdot$atom). Transitions to the trilayer film [Figure \ref{fig2}.~(b)] mark a critical turning point. The emergence of interlayer coupling and the expanded periodicity along the $z$-direction modify the confinement potential, enhancing the overlap between $s$, $p$, and $d$ orbitals. Concurrently, the increased number of atomic layers introduces additional quantization channels. These factors conspire to reconstruct the electronic landscape, substantially boosting $N(\varepsilon_\text{F})$ to a maximum of $0.25$ states/(eV$\cdot$atom). Upon further thickening to five layers [Figure \ref{fig2}.~(c)], the weakening of quantum confinement manifests as a dense manifold of subbands. Compared to thinner films, the energy bands become more closely spaced and their discrete nature begins to blur. By the septuple-layer limit (7L) [Figure \ref{fig2}.~(d)], the electronic structure---including band dispersion and DOS features---asymptotically converges to that of bulk Cu, with $N(\varepsilon_\text{F})$ stabilizing at around $0.24$ states/(eV$\cdot$atom).

\begin{figure*}[ht]
\centering
\includegraphics[width=1\textwidth]{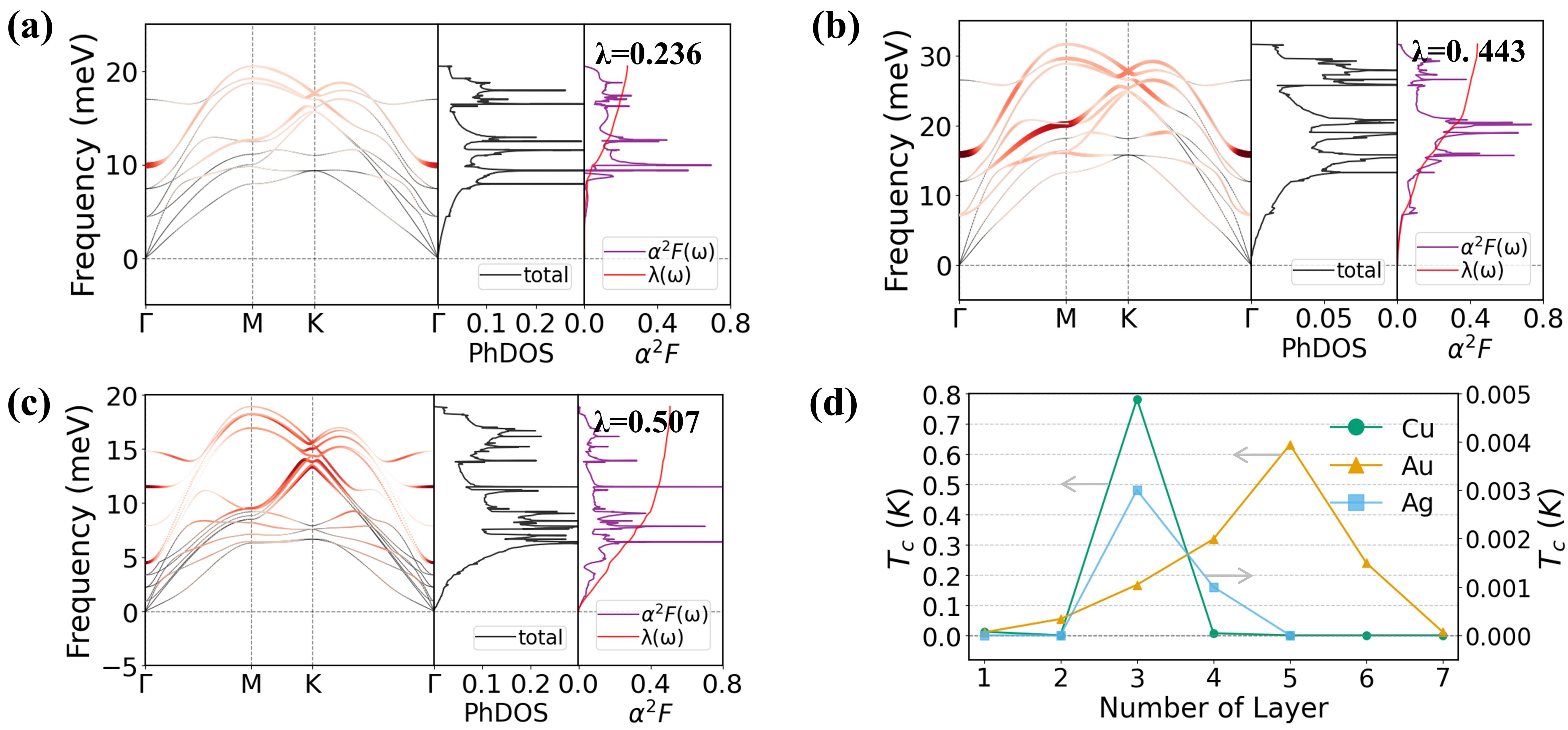}
\caption{Phonon properties and superconductivity. 
(a--c) Phonon and EPC properties of trilayer Ag, trilayer Cu, and pentalayer Au thin films. For each: EPC-weighted phonon dispersion (red, scaled by $\lambda_{\bm{q}\nu}$), phonon density of states (PhDOS), Eliashberg function $\alpha^2 F(\omega)$, and cumulative EPC $\lambda(\omega)$. (d) Superconducting transition temperature $T_\text{C}$ versus layer number for Au, Ag, and Cu, plotted with distinct y-axis scales.}
\label{fig3}
\end{figure*}

\subsubsection{\textbf{Phononic and superconducting properties}}
With the electronic structure evolution established, we now turn to the lattice dynamics and superconductivity of Au, Ag, and Cu thin films. All optimized structures are dynamically stable, as confirmed by the absence of imaginary frequencies in their phonon spectra [Figure~S4].~This dynamical stability serves as the prerequisite for our subsequent investigation of their superconducting properties.

\begin{table}[htbp]
\centering
\renewcommand{\arraystretch}{1.2}
\caption{Superconducting properties of Ag, Cu, Au thin flms with diferent thicknesses.}
\begin{ruledtabular}
\begin{tabular}{cccccc}
& Layer & $T_{\text{C}}$ (K) & $\lambda$ & $\omega_{\text{log}}$ (meV) & $N(E_F)$ [states/(eV$\cdot$atom)] \\
\midrule
\multirow{5}{*}{\centering Ag} & 1L & 0.00 & 0.19 & 174.45 & 0.14 \\
& 2L & 0.00 & 0.24 & 53.20 & 0.17 \\
& 3L & 0.003 & 0.24 & 146.01 & 0.18 \\
& 4L & 0.001 & 0.23& 80.24 & 0.18 \\
& 5L & 0.00 & 0.19 & 106.83 & 0.18 \\
\midrule
\multirow{7}{*}{\centering Cu} & 1L & 0.01 & 0.28 & 115.02 & 0.19 \\
& 2L & 0.00 & 0.22 & 70.79 & 0.24 \\
& 3L & 0.78 & 0.44 & 132.88 & 0.25 \\
& 4L & 0.01 & 0.25 & 196.41 & 0.25 \\
& 5L & 0.00 & 0.13 & 181.53 & 0.24 \\
& 6L & 0.00 & 0.21 & 183.69 & 0.24 \\
& 7L & 0.00 & 0.23 & 138.40 & 0.24 \\
\midrule
\multirow{7}{*}{\centering Au} & 1L & 0.01 & 0.27 & 86.17 & 0.17 \\
& 2L & 0.05 & 0.34 & 55.01 & 0.20 \\
& 3L & 0.17 & 0.41 & 42.17 & 0.25 \\
& 4L & 0.32 & 0.43 & 61.82 & 0.25 \\
& 5L & 0.63 & 0.51 & 57.05 & 0.23 \\
& 6L & 0.24 & 0.39 & 79.89 & 0.23 \\
& 7L & 0.01 & 0.29 & 70.14 & 0.23 \\
\label{table2}
\end{tabular}
\end{ruledtabular}
\end{table}

Among the noble metal films, Ag exhibits the weakest superconducting response. Its maximum $T_\text{C}$ reaches only $\sim$3 mK in the trilayer structure [Table \ref{table1}]. This marginal behavior arises from its intrinsically weak EPC ($\lambda < 0.3$, peaking at 0.24), which is constrained by both electronic and phononic bottlenecks. Electronically, $N(\varepsilon_\text{F})$ remains low and only weakly dependent on thickness. Meanwhile, the Ag $d$-band center is located $\sim$3 eV below $\varepsilon_\text{F}$, which limits the effectiveness of quantum confinement in enhancing $N(\varepsilon_\text{F})$ [Figure S2]. Consequently, $N(\varepsilon_\text{F})$ remains low at $\sim$0.17 states/(eV$\cdot$atom) [Table \ref{table2}]. In addition, Ag presents a notably stiff lattice. Even in the trilayer limit, its logarithmic average phonon frequency $(\omega_{\rm log})$ remains high at 146 K, substantially exceeding values for comparable Au and Cu films. The phonon density of states (PhDOS) is concentrated in the 10--20 meV range, where the Eliashberg spectral function $\alpha^2F(\omega)$ appears broad and flat with low intensity [Figure \ref{fig3}.~(a)], indicating inefficient coupling across the entire frequency spectrum. This decoupling is further evidenced by the uncorrelated evolution of $\lambda$ and $N(\varepsilon_\text{F})$ with thickness [Figure \ref{fig2}.~(e)]. These results underscore a fundamental lesson: low-dimensional superconductivity cannot be achieved by optimizing a single parameter; it demands a synergistic interplay between electronic density and lattice softness.

In stark contrast to Ag, Cu thin films exhibit remarkable enhanced superconductivity. Both $T_\text{C}$ and $\lambda$ peak in the 3L-Cu(111) structure, attaining $0.78\ \text{K}$ and $0.44$, respectively [Figure \ref{fig2}.~(e) and Table \ref{table2}].We attributed this enhancement to two cooperative mechanisms. First, quantum confinement boosts $N(\varepsilon_\text{F})$ to $0.25\ \text{states/(eV$\cdot$atom)}$---the highest among all studied thicknesses---thereby maximizing the phase space for Cooper pairing. Second, spectral analysis of $\alpha^2 F(\omega)$ reveals a layer-selective activation of coupling modes: strong EPC is concentrated in the mid-to-high frequency range ($10$--$25\ \text{meV}$), forming a distinct peak. This frequency selectivity is rooted to Cu's lighter atomic mass, which results in stiffer bonds and higher vibrational frequencies, allowing optical-like modes to dominate pairing. Together, these mechanisms produce the pronounced $\lambda$ in 3L-Cu(111), which is closely associated with the increased $N(\varepsilon_\text{F})$ and enhances the electron--phonon pairing interaction. Consequently, the peak in $T_\text{C}$ observed in the trilayer limit [Figure \ref{fig3}.~(d)] can be viewed as the result of the simultaneous optimization of the electronic density and effective coupling to high-frequency phonons.

Conversely, the superconducting enhancement in Au thin films is governed primarily by phonon softening. As illustrated in Figure \ref{fig2}.~(e), $\lambda$ and $\omega_{\rm log}$ exhibit a clear inverse correlation, implying the existence of an optimal softening window. As detailed in Table \ref{table2}, both $T_\text{C}$ ($0.63\ \text{K}$) and $\lambda$ ($0.51$) peak simultaneously in the 5L-Au(111) structure, where $\omega_{\rm log}$ softens to $57.05\ \text{K}$ [Figure \ref{fig3}.~(c)]. This optimal window is bounded by two detrimental regimes. On the ``over-softened'' side, 3L-Au (111) exhibits an excessively low $\omega_{\rm log}$ of $42.17\ \text{K}$. Despite possessing a high $N(\varepsilon_\text{F})$ (0.25 states/(eV$\cdot$atom)), its coupling strength ($\lambda = 0.41$) remains suppressed, which is primarily attributed to the low contribution from the electron–phonon coupling matrix element $\lvert g_{nm,v=8}(\boldsymbol{k},\mathbf{0}) \rvert$ for the $8 {\textit{th}}$ phonon mode at the $\Gamma$ point [Figure S5]. On the ``hardened''side, increasing thickness to 6L-Au (111) raises $\omega_{\rm log}$ to $79.89\ \text{K}$, causing a sharp decline in both $\lambda$ and $T_\text{C}$. The stability of $N(\varepsilon_\text{F})$ throughout this range ($\sim0.23\ \text{states/(eV$\cdot$atom)}$) effectively rules out electronic states as the driving factor. These findings collectively demonstrate that superconductivity in Au is tuned by the degree of lattice softening, with an optimal $\omega_{\rm log}$ range ($\sim55$--$65\ \text{K}$) maximizing the coupling efficiency.

The distinct superconducting responses of Ag, Cu, and Au ultrathin films provide a rigorous testbed for reassessing prevailing theories. We benchmark our first-principles results against the pioneering analytical model by Ummarino and Zaccone (U\&Z) \cite{re31, re32}, which incorporated quantum confinement into Eliashberg theory. The U\&Z model predicted a $T_\text{C}$ peak at $\sim 3$ atomic layers (driven by confinement-enhanced $N(\varepsilon_\text{F})$, with a magnitude ordering of Au ($1.04\ \text{K}$) $>$ Ag ($0.29\ \text{K}$) $>$ Cu ($0.12\ \text{K}$)). While our calculations confirm the qualitative prediction of a non-monotonic $T_\text{C}$ evolution [Figure \ref{fig2}.~(e) and Figure \ref{fig3}.~(d)], we observe substantial quantitative deviations, specifically in the hierarchy of Cu ($0.78\ \text{K}$) $>$ Au ($0.63\ \text{K}$) $>$ Ag ($\sim 3\ \text{mK}$). It is important to note that while the U\&Z framework is conceptually robust (even extending to amorphous systems), its numerical predictions for crystalline films depend critically on input parameters. Our analysis identifies two key sources of discrepancy stemming from the model's simplified treatment of atomic-scale processes:

First, regarding structural relaxation: The U\&Z model employs a simplified geometric scaling, deriving electronic properties solely from film thickness and bulk parameters via scaling factors. In contrast, our calculations [Table \ref{table1}] reveal that realistic Au and Ag films undergo complex internal relaxation, including significant surface contraction and non-uniform interlayer spacing variations. This atomic-scale renormalization, which fundamentally alters the electronic boundary conditions, is absent in the U\&Z formalism.

Second, regarding phonon physics: The U\&Z model adopts a fixed bulk $\alpha^2 F(\omega)$ \cite{re63}, implicitly assuming that the phonon spectrum and coupling strength are independent of film thickness. However, our results [Figure~S4] demonstrate a profound thickness dependence of the phonon landscape. As the film thins, the spectrum transitions from a bulk-like continuum to discrete branches dominated by surface modes. Specifically, in the monolayer and bilayer limits, localized surface modes dominate the low-energy spectrum, with their dispersion near the $\Gamma$ point becoming notably gentle. This spectral reshaping directly dictates the coupling strength $\lambda$. By neglecting this evolution, the fixed-spectrum approximation fails to capture the material-specific interplay between confinement and lattice dynamics (e.g., the phonon softening in Au\ vs.\ stiffening in Cu).

While the U\&Z model significantly simplifies structural relaxation in Au and Ag, the Cu system serves as an ideal testbed for assessing its performance. Given the minimal variation of Cu's lattice parameters with thickness—featuring a lattice constant of $\sim2.54$~\AA{} and an interlayer spacing of $\sim2.10$~\AA{}—the model's geometric scaling assumption is particularly appropriate in this case [Table \ref{table1}]. Within the U\&Z framework, $\lambda$ and $N(\varepsilon_\text{F})$ for ultrathin Cu films are estimated at 0.24 and 0.15~states/(eV$\cdot$atom), compared to 0.44 and 0.25~states/(eV$\cdot$atom) from our first-principles calculations. The qualitative consistency suggests that the model captures the basic trend of confinement-driven enhancement. However, even for Cu—where the structural approximation holds best—the reliance on a fixed bulk phonon spectrum limits the model, yielding a predicted $T_\text{C}$ of only $0.12$~K. This stands in sharp contrast to our first-principles results (EPW: $1.20$~K; QE: $0.78$~K). This substantial quantitative discrepancy underscores that an accurate theoretical description of $T_\text{C}$ in ultrathin films necessitates the explicit incorporation of thickness-dependent phonon renormalization.

In summary, our work elucidates that low-dimensional superconductivity is not determined by a universal confinement rule alone, but emerges from the complex, element-specific interplay of quantum well states and phonon renormalization.

\subsection{Heterointerface Engineering}
\subsubsection{\textbf{h-BN/3L-Cu(111) heterostructure}}
With a foundational understanding of the intrinsic superconducting mechanisms established, we now explore the potential of interface engineering to functionalize these noble metal films. Our stability analysis reveals that a trilayer of Au or Ag on h-BN is dynamically unstable, exhibiting imaginary phonon modes [Figure~S6]. Even a 30$^\circ$ rotation reduces the strain from $\sim$18\% to $\sim$0.2\% (Au) and $\sim$0.054\% (Ag), but both interfaces remain dynamically unstable due to long wavelength phonon instabilities [Note S2 and Table S1]. In contrast, Cu-based systems are robust, a finding corroborated by the prevalent experimental use of Cu foils as epitaxial substrates for h-BN and graphene growth \cite{re45,re66}. We therefore focus on the 3L-Cu(111) film, which exhibits superior superconducting performance, as the host system. Using the dynamically stable h-BN/3L-Cu(111) heterostructure as a central model, we systematically evaluate four high-symmetry stacking configurations (B-top, N-top, hollow, and bridge) [Figure S7 and Table S2]. The N-top configuration is the global ground state with formation energy of $-0.04$ eV/f.u., followed by B-top at $-0.03$ eV/f.u., and both bridge and hollow at $-0.02$ eV/f.u. above N-top. Accordingly, we focus on the N-bonded and B-bonded interfaces as the two most relevant cases for the subsequent analysis.

\begin{figure*}
\centering
\includegraphics[width=1\textwidth]{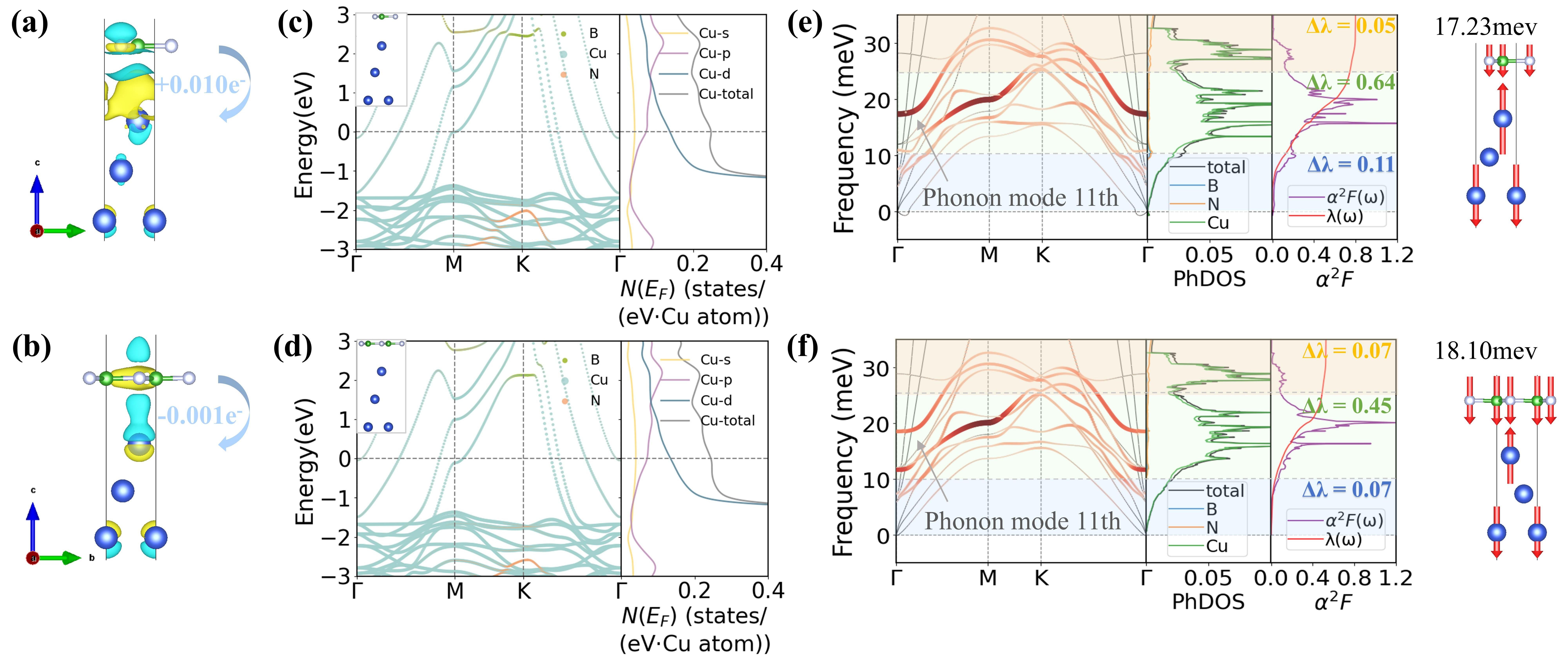} 
\caption{Interface-configuration-tuned electronic and EPC properties in h-BN/3L-Cu(111) heterostructure. (a, b) Differential charge density plots for B-site and N-site interfaces in side view. (c, d) Orbital-projected band structures for B-site and N-site, respectively. (e, f) Corresponding EPC properties: $\mathrm{\lambda}_{qv}$-scaled phonon dispersions (red), phonon density of states, Eliashberg function $\alpha^2 F(\omega)$, and cumulative $\lambda(\omega)$.The rightmost panels in (e) and (f) show the vibrational eigenmodes associated with the 11 $th$ phonon branch.}
\label{fig4}
\end{figure*}

\begin{ruledtabular}
\renewcommand\arraystretch{1.2}
\begin{table*}
\caption{Superconducting properties and electronic structure parameters, and charge transfer for h-BN/3L-Cu(111) heterostructure.}  
\begin{tabular}{ccccccccc}
\multicolumn{2}{c}{} & Optimized(\AA) & Interlayer spacing (\AA) & $T_\text{C}$(K) & $\lambda$ & $\omega_{log}$(K) & $N(\varepsilon_\text{F})$ states/(eV$\cdot$atom)  & $\Delta Q$\\ 
\midrule 
\multirow{2}{*}{monolayer} & B-site & 2.50 & 3.10 & 7.00 & 0.80 & 158.42 & 0.24 & +0.01e \\
                          & N-site & 2.51 & 2.94 & 3.23 & 0.59 & 170.58 & 0.24 & -0.001e 
\label{table3}
\end{tabular}
\end{table*}
\end{ruledtabular}

\begin{figure*}[htbp]
  \centering
  \includegraphics[width=0.9\textwidth]{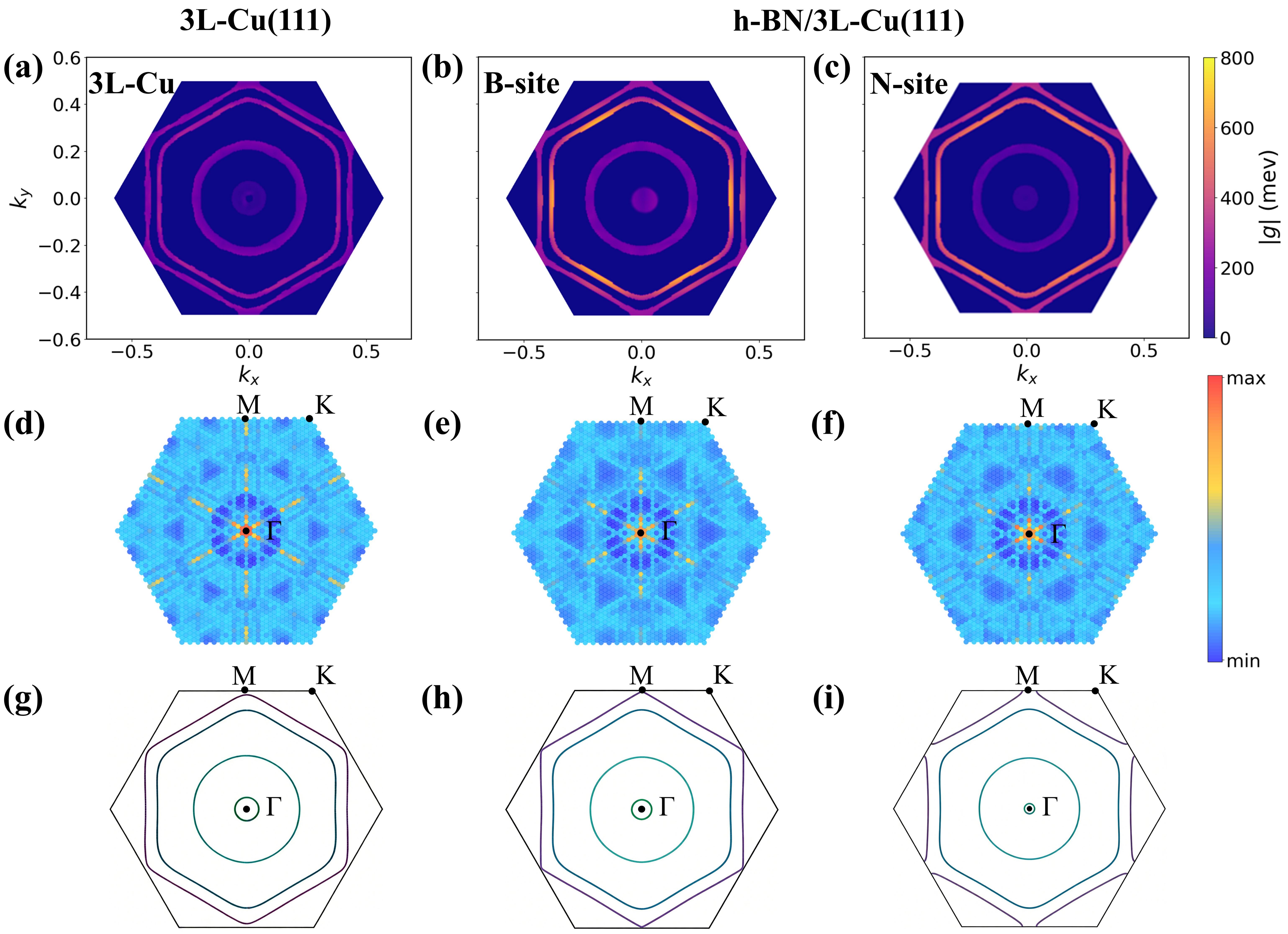} 
  \caption{\textit{k}-dependent electron--phonon coupling matrix element, Fermi surface nesting function, and Fermi surface.
  (a) Electron--phonon coupling matrix element $\lvert g_{nm,v=8}(\boldsymbol{k},\mathbf{0}) \rvert$ for the $8 {\textit{th}}$ phonon mode at the $\Gamma$ point in the 3L-Cu(111) system.
  (b, c) Electron--phonon coupling matrix element $\lvert g_{nm,v=11}(\boldsymbol{k},\mathbf{0}) \rvert$ for the $11 {\textit{th}}$ phonon mode at the $\Gamma$ point in the h-BN/3L-Cu(111) heterostructure (b: B-site interface; c: N-site interface).
  The color scale in this region represents the magnitude of the matrix element $g$ (in meV).
  (d) Fermi surface nesting function for the 3L-Cu(111) system.
  (e, f) Fermi surface nesting function for the h-BN/3L-Cu(111) heterostructure (e: B-site interface; f: N-site interface).
  (g--i) Fermi surface for the 3L-Cu(111) system and the h-BN/3L-Cu(111) heterostructure (h: B-site interface; i: N-site interface).}
  \label{fig5}
\end{figure*}

\begin{ruledtabular}
\renewcommand\arraystretch{1.2}
\begin{table*}
\caption{Superconducting properties and electronic structure parameters, and charge transfer for h-BN/3L-Cu(111) heterostructure.}  
\begin{tabular}{cccccccccc}
\multicolumn{2}{c}{} & staking & Optimized(\AA) & Interlayer spacing (\AA) & $T_\text{C}$(K) & $\lambda$ & $\omega_{log}$(K) & $N(\varepsilon_\text{F})$ states/(eV$\cdot$atom)  & $\Delta Q$ \\
\midrule
\multirow{4}{*}{Bilayer} & \multirow{2}{*}{B-site} & AB & 2.51 & 3.08/3.09 & 7.42 & 0.86 & 146.54 & 0.25 & +0.02e \\
& & BA & 2.51 & 3.07/3.09 & 7.55 & 0.86 & 148.64 & 0.25 & -0.02e \\
\cmidrule{2-10}
& \multirow{2}{*}{N-site} & AB & 2.51 & 3.08/2.92 & 2.15 & 0.50 & 200.58 & 0.24 & +0.02e \\
& & BA & 2.51 & 3.06/2.94 & 2.28 & 0.51 & 201.41 & 0.24 & -0.02e \\
\midrule
\multirow{2}{*}{sandwich} & B-site & -- & 2.51 & 3.10/2.08/2.95 & 1.80 & 0.48 & 200.64 & 0.24 & -0.21e/-0.01e \\
& N-site & -- & 2.51 & 2.95/2.06/3.01 & 1.29 & 0.45 & 210.63 & 0.24 & -0.01e/-0.01e 
\label{table4}
\end{tabular}
\end{table*}
\end{ruledtabular}

In the h-BN/3L-Cu(111) heterostructure, the introduction of a monolayer h-BN cover significantly enhances the $T_\text{C}$ of the underlying copper film by approximately 4 to 9 times [Table \ref{table3}], rather than merely serving as a passivation layer. Crucially, this enhancement exhibits a pronounced dependence on the specific type of interfacial atom: whether the Cu atom sits below a Boron (B) or Nitrogen (N) site. To decode the mechanism driving this site-specific modulation, we first focused on potential dominant factors: charge transfer and $N(\varepsilon_\text{F})$. Bader charge analysis indicates negligible interfacial charge transfer (B site: $\Delta Q = +0.01\ \text{e}$; N site: $\Delta Q = -0.001\ \text{e}$) [Figures \ref{fig4}.~(a, b)]. Similarly, the $N(\varepsilon_\text{F})$ varies by less than 5\% across configurations (intrinsic Cu: $0.25$; B-interface: $0.24$; N-interface: $0.24\ \text{states/(eV·atom)}$) [Figures \ref{fig4}.~(c, d)]. These minimal variations confirm that the dramatic $T_\text{C}$ enhancement cannot be ascribed to simple doping or DOS effects.

In marked contrast to the nearly constant $N(\varepsilon_\text{F})$, the EPC constant $\lambda$ exhibits strong configuration dependence across interfacial arrangements. Notably, $\lambda$ reaches $0.80$ for the B-site configuration, significantly higher than for the N-site ($0.56$) and the pristine three-layer Cu film ($0.44$). Spectral analysis of $\lambda(\omega)$ [Figures \ref{fig4}.~(e, f); full phonon spectrum in Figure.~S9] indicates that the B-site interface enhances the contribution from mid-frequency phonons ($10$–$25\ \text{meV}$), which account for $\sim 80\%$ of the total $\lambda$ ($0.64$), compared with $76\%$ ($0.52$) for the N-site interface and $68\%$ ($0.30$) for the intrinsic Cu film.

The microscopic origin of this enhancement is further elucidated by analyzing the momentum dependence of the EPC matrix elements, $g_{nm,\nu}(\boldsymbol{k},\boldsymbol{q})$, and the role of Fermi surface nesting (FSN). To pinpoint the critical scattering channels, we first examined the phonon linewidths ($\gamma_{\boldsymbol{q}\nu}$) [Figures \ref{fig4}.~(e, f)]. This analysis reveals that the disparity between the B- and N-site interfaces is concentrated in the mid-frequency phonon branches near $\Gamma$---specifically, the 11$th$ mode of the heterostructures, which corresponds topologically to the 8$th$ mode of pristine 3L-Cu(111) [Figure~S10]. Guided by this mode-specific sensitivity, we mapped the ${k}$-space distribution of the coupling strength $\left|g_{nm,\nu}(\boldsymbol{k},0)\right|$ for this critical mode near $\Gamma$ [Figures \ref{fig5}.~(a-c)].

The momentum-space distributions are strongly interface-dependent, showing localization in annular regions near the Fermi surface, consistent with scattering dominated by Fermi-surface states. For the B-site interface, $\left|g_{nm,v}(\boldsymbol{k},0)\right|$ displays pronounced, highly localized bright spots along the edges of the hexagonal Brillouin zone [Figure \ref{fig5}.~(h)], particularly near the M--K directions, signaling a marked coupling enhancement in these restricted momentum regions. In sharp contrast, while the N-site interface also shows an overall increase, its enhancement is distributed more uniformly along the zone edges, with far less pronounced spectral features. Importantly, the FSN function exhibits moderate features along the $\Gamma$--M direction for both interfaces and is virtually identical for the two interfaces, showing no new nesting vectors and only minimal configurational differences [Figure \ref{fig5}.~(e, f)]. Consequently, the associated electronic phase space remains essentially unchanged, effectively excluding FSN as the primary cause of the enhanced coupling at the B-site. Instead, the enhancement is primarily attributed to a strong, momentum-region-specific amplification of the EPC matrix elements.

The microscopic origin of this interface modulation is further elucidated by analyzing the interplay between specific out-of-plane phonon modes and Fermi surface topology evolution. From lattice dynamics, the h-BN overlayer exerts mechanical constraints that hinder EPC enhancement. As the interface shifts from pristine Cu to the B- and N-site configurations, the out-of-plane phonon frequency hardens ($15.75$, $17.23$, $18.10\ \text{meV}$) [Figures \ref{fig4}.~(e, f) and Figure~S10], while the out-of-plane vibrational amplitude of surface Cu atoms decreases ($u_z$: $0.71$, $0.65$, $0.55$) derived from vibration eigenvector. The N-site interface exhibits the strongest clamping effect, yielding the stiffest lattice and smallest amplitude. According to the canonical scaling $\lambda \propto 1/\omega^2$, these trends imply a reduction in coupling upon functionalization, contradicting the computed significant enhancement. The critical factor enabling enhancement is the interface-induced Fermi surface Lifshitz transition: intrinsic 3L-Cu has a closed hexagonal Fermi surface well-separated from Brillouin zone boundaries [Figure \ref{fig5}.~(g)]. At the B-site interface, the Fermi surface expands to become precisely tangent to the M point [Figure \ref{fig5}.~(h)]. This critical topological feature populates high-susceptibility electronic states near the Fermi level ($E_F$), enabling efficient scattering. In contrast, the N-site configuration's Fermi surface expands further, breaking its closure and moving away from this critical tangency [Figure \ref{fig5}.~(i)].

Consequently, the B-site's superior EPC arises from a synergy where electronic enhancement outweighs lattice stiffening \cite{re69}. While the B-site lattice is stiffer than the intrinsic film, its relatively large vibrational amplitude ($0.65$) still couples robustly with M-point critical electronic states (induced by the tangent Fermi surface topology) to generate substantial scattering matrix elements, maximizing $\lambda$. The N-site, conversely, suffers from excessive lattice stiffening and a significantly reduced vibrational amplitude ($0.55$), limiting its ability to modulate electronic states. The intrinsic 3L-Cu(111), lacking favorable Fermi surface topology and interfacial electronic perturbation, exhibits the weakest coupling despite its softest lattice and largest amplitude.

\begin{figure}
  \centering
  \includegraphics[width=0.5 \textwidth]{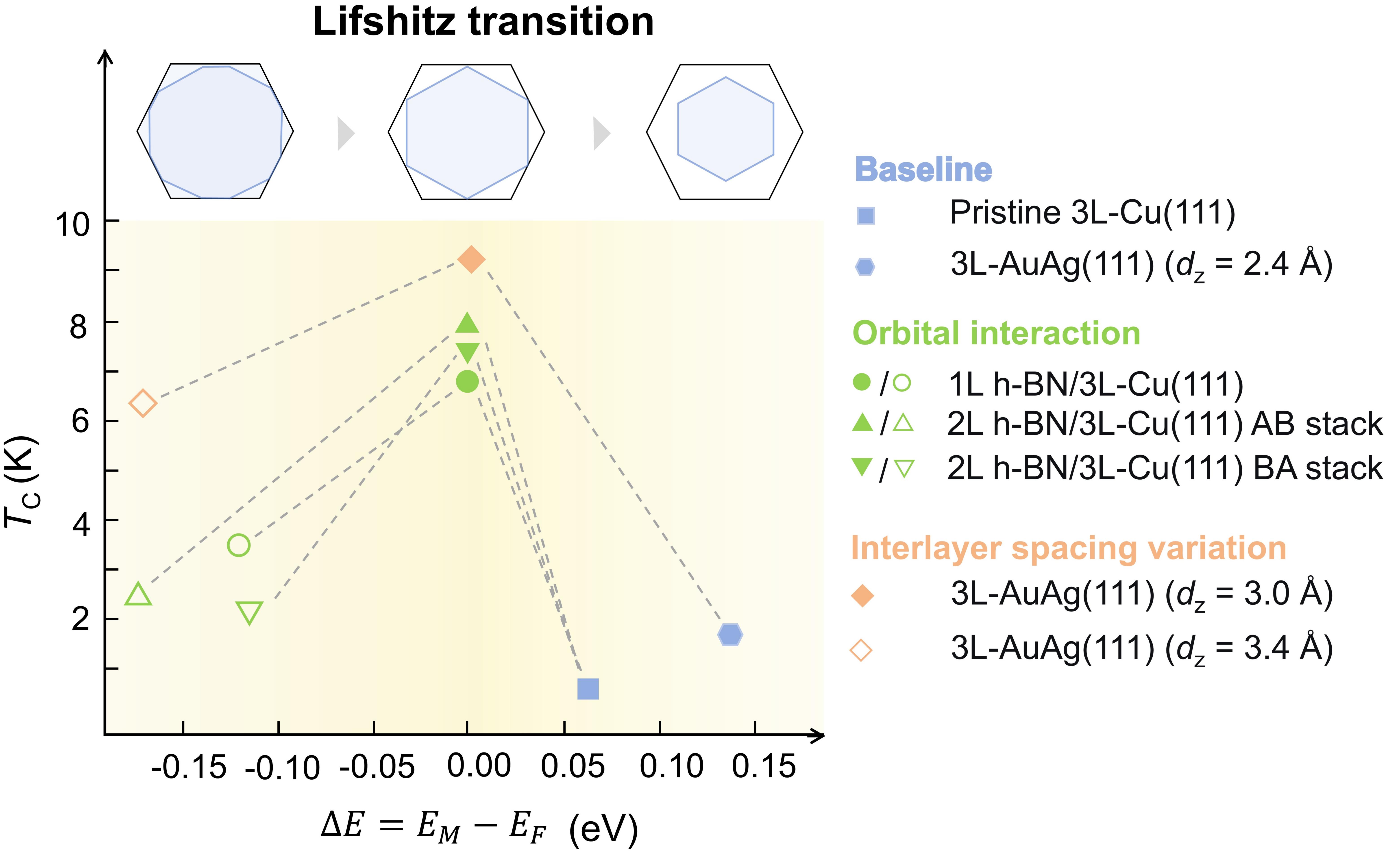} 
  \caption{Dependence of $T_C$ on Fermi surface position across the Lifshitz transition. 
$\Delta E = E_M - E_F$ measures the energetic distance from the Fermi level to the band edge at the Brillouin zone boundary at the M point.
Green/orange symbols indicate orbital interaction / interlayer spacing ($d_z$) variation; filled/open symbols denote B‑/N‑site contributions (or their equivalents in AuAg).
For the 3L‑Au/Ag(111) system, the interlayer spacings are $d_z = 2.4$~\AA{} (N‑site equiv.), $d_z = 3.0$~\AA{} (B‑site equiv.), and $d_z = 3.4$~\AA{} (pristine 3L‑Cu(111) equiv.).
Data are shown for 1L‑h‑BN/3L‑Cu(111), 2L‑h‑BN/3L‑Cu(111) (AB/BA stacking), and the above AuAg spacings. Note that $T_C$ values for the AuAg systems are scaled by a factor of 10 ($T_C \times 10$) for visual clarity on the same axis.
In all systems, $T_C$ is significantly enhanced as the Fermi surface approaches the zone boundary (Lifshitz transition point).}
  \label{fig6}
\end{figure}

The universality of this Lifshitz-induced strategy is further confirmed in a bimetallic system. The [111]-oriented Au/Ag heterostructure (3L-Au(111) on 3L-Ag(111))~\cite{re70} is dynamically stable and exhibits a Fermi surface topology analogous to the N-site interface in our h-BN/Cu system [Figure S12]. By artificially tuning the vertical spacing between the bottommost Au and topmost Ag layers, we can drive the Fermi surface either toward the Brillouin-zone boundary (mimicking the B-site interface near the Lifshitz transition) or away from it (mimicking pristine 3L-Cu). Increasing the spacing, however, destabilises the structure and produces imaginary phonon modes. To assess the EPC trend associated with the modified Fermi surface, we computed EPC and $T_\mathrm{C}$ using only the real-frequency phonon modes. The $T_\mathrm{C}$ serves as a theoretical benchmark --- the maximum enhancement possible if the favourable Fermi surface could be stabilised. As shown in Table S4, the EPC evolution with interlayer spacing in Au/Ag closely reproduces the Fermi-surface topological transitions observed in h-BN/3L-Cu(111). The EPC is significantly enhanced when the Fermi surface approaches the zone boundary, albeit at the cost of dynamical instability. These results support the universality of the Lifshitz-transition-driven enhancement mechanism while highlighting the need for structural stabilisation in practical systems [Figure \ref{fig6}].

\subsubsection{\textbf{h-BN/h-BN/3L-Cu(111) and h-BN/3L-Cu(111)/h-BN heterostructures}}
To validate the universality of this mechanism and explore complex interface designs, we further constructed a bilayer h-BN/h-BN/3L-Cu(111) heterostructure with switchable slip-stacking orders (AB/BA). This setup allows us to disentangle the effects of interfacial chemistry from stacking ferroelectricity. Remarkably, our results [Table \ref{table4}] indicate that stacking order (AB vs. BA) plays a negligible role, altering $T_\text{C}$ by only $\sim 0.1\ \text{K}$. The decisive factor remains the immediate interfacial atomic species. The B-site interface consistently delivers superior performance, achieving a high $T_\text{C}$ of $\sim 7.49\ \text{K}$ and $\lambda \approx 0.86$, compared to $\sim 2.22\ \text{K}$ and $\lambda \approx 0.51$ for the N-site interface. A critical insight from this bilayer system is the decoupling of $T_\text{C}$ from $N(\varepsilon_\text{F})$: despite nearly identical DOS values at B- and N-interfaces (difference $<0.4\%$), a $\sim 5\ \text{K}$ difference in $T_\text{C}$ persists [Table \ref{table4}]. This finding provides compelling evidence that interface-orbital-specific coupling, rather than DOS magnitude, is the dominant order parameter.

Further analysis confirms that the bilayer h-BN covered 3L-Cu(111) inherits the EPC enhancement mechanism of the monolayer [Figures~S13, S14 and Table \ref{table4}]. The B-site interface maintains a symmetric, covalent-like bonding environment with minimal charge transfer ($\Delta Q \approx \pm0.02\ \text{e}$) and softer lattice dynamics ($\omega_{\rm log} = 146.54\ \text{K}$). Conversely, the N-site interface suffers from localized electrostatic stiffening ($\omega_{\rm log} \approx 200.10\ \text{K}$). Relative to the monolayer configuration (h-BN/3L-Cu (111)), the slip-stacked bilayer enhances the cooperativity of these orbital interactions without compromising lattice integrity, demonstrating that this design strategy is robust against structural complexity.        

The consistent success of B-site engineering in monolayer and bilayer systems raises a fundamental question: Is ``more'' always ``better''? To answer this, we examine the extreme case---a symmetric h-BN/3L-Cu(111)/h-BN sandwich architecture---which maximizes interfacial confinement [Figure \ref{fig1}~(e), Figure S17 and Table S5].~Counterintuitively, our calculations reveal that this dual-interface confinement drastically suppresses $T_\text{C}$ to below $2$~K (Note~S3). While one might expect additive coupling channels, the rigid physical confinement imposes severe phonon hardening across the entire film. More crucially, the electronic structure analysis [Figure S17] shows that the bands near the M-point shift further away from the Fermi level, deviating from the critical Lifshitz transition point observed in the asymmetric B-site h-BN/3L-Cu(111) system. This combined effect of lattice stiffening and unfavorable electronic topology outweighs the benefits of increased interfacial area \cite{re71, re72}. This pivotal finding delineates the operational boundary of interface engineering: the path to high $T_\text{C}$ in noble metals lies in moderate, asymmetric interface coupling that balances optimal orbital hybridization with lattice flexibility, rather than maximal symmetric confinement.

\section{CONCLUSIONS}
In summary, this first-principles study establishes a unified physical framework for superconductivity, bridging the gap from intrinsic few-layer noble metals to engineered heterostructures. Our calculations elucidate the material-specific nature of low-dimensional superconductivity: it is fundamentally limited in trilayer Ag, DOS-driven in trilayer Cu, and phonon-softening-mediated in pentalayer Au.~Beyond these intrinsic behaviors, we propose specific interfacial stacking configuration as a robust mechanism for active control. We demonstrate that the selection of the interfacial bonding atom (B vs.\ N) effectively induces a synergistic interplay between favorable Fermi surface topology and strong EPC, thereby modulating the superconducting state. This approach aligns with recent experimental breakthroughs in 2D semiconductor contacts, where metallic bonding was shown to improve lattice coherence and enhance electronic coupling \cite{re73}. Specifically, we identify the B-site configuration---characterized by its induction of a critical interface-driven Lifshitz transition and concerted action with a relatively preserved lattice vibrational response---as a key design principle for the synergistic enhancement of EPC. Furthermore, our comparative study of slip-stacked configurations provides compelling evidence that EPC enhancement, rather than the electronic DOS alone, governs superconducting performance. Finally, insights from the sandwich heterostructure delineate the boundary of this enhancement: while the $T_\text{C}$ of the double-interface structure exceeds that of intrinsic Cu, it is notably suppressed compared to single-interface heterostructures due to the deviation from the critical Lifshitz point and phonon hardening imposed by excessive confinement. This pivotal finding establishes that moderate, optimized interface coupling---balancing electronic enhancement against lattice hardening---is the essential design criterion. Collectively, these principles outline a general strategy for functionalizing noble metal superconductors and open avenues for future research, including experimental realization, extension to other material families, and dynamic control in quantum devices.

\begin{acknowledgments}
This work was financially supported by the National Natural Science Foundation of China (Grants 12304165 and 12304086), the Natural Science Foundation of Inner Mongolia Autonomous Region (Grant 2023QN01003), the “Grassland Talents” project of the Inner Mongolia autonomous region (Grant 21200-242920), and the Startup Project of Inner Mongolia University (Grant 21200-5223733). 
\end{acknowledgments}

\clearpage
\bibliography{reference}
\end{document}